\documentclass[11pt]{article}

\usepackage{fullpage}
\usepackage{amsmath,amsfonts,amsthm,mathrsfs,xspace,hyperref,graphicx,relsize,bm}
\usepackage{mathpazo}

\usepackage{float}
\usepackage{endnotes}
\usepackage{enumitem}
\usepackage{color}
\usepackage{amssymb,latexsym}
\usepackage{commath}

\theoremstyle{plain}
\newtheorem{thm}{Theorem}
\newtheorem{theorem}[thm]{Theorem}

\newtheorem{proposition}[thm]{Proposition}

\newcommand{\ket}[1]{|#1 \rangle}
\newcommand{\bra}[1]{\langle #1|}

\newcommand{\braket}[2]{\langle #1|#2 \rangle}
\newcommand{\ketbra}[2]{|#1 \rangle\!\langle #2 |}
\newcommand{\Tr}{\mathrm{Tr}}
\newcommand{\Id}{\mathbb{I}}

\theoremstyle{definition}

\DeclareMathOperator*{\Ex}{\mathbb{E}}

\newcommand{\Hilb}{\mathcal{H}}
\newcommand{\C}{\mathbb{C}}

\newcommand{\setft}[1]{\mathrm{#1}}
\newcommand{\State}{\setft{S}}
\newcommand{\Density}{\setft{D}}

\begin{document}

\title{A simple proof of Renner's exponential de Finetti theorem}
\author{Thomas Vidick\thanks{Department of Computing and Mathematical Sciences, California Institute of Technology. Supported by NSF CAREER Grant CCF-1553477, an AFOSR YIP award, and the IQIM, an NSF Physics Frontiers Center (NFS Grant PHY-1125565) with support of the Gordon and Betty Moore Foundation (GBMF-12500028). Email: {\tt vidick@cms.caltech.edu}} \and Henry Yuen\thanks{UC Berkeley. Partially supported by Simons Foundation grant 360893 and National Science Foundation
grant 1218547. Email: {\tt hyuen@csail.mit.edu}}}
\date{}
\maketitle

\begin{abstract}
	We give a simple proof of the exponential de Finetti theorem due to Renner. Like Renner's proof, ours combines the post-selection de Finetti theorem, the Gentle Measurement lemma, and the Chernoff bound, but avoids virtually all calculations, including any use of the theory of types.
	\end{abstract}

\section{Introduction}

In quantum information theory a \emph{de Finetti theorem} expresses the fact that a quantum state $\rho \in (\C^d)^{\otimes n}$ that is invariant under permutation of its $n$  subsystems is close to a mixture of tensor product states. The first such result is due to de Finetti~\cite{de1937prevision}, who showed that any classical distribution that is \emph{infinitely exchangeable} can be expressed as a convex combination of product distributions. Many variants of de Finetti theorems have since been shown, and we will not attempt to survey them here. They have found applications to quantum information theory~\cite{duan2015zero}, cryptography~\cite{renner2008security}, and complexity~\cite{lancien2016flexible}, among others. 

The simplest quantitatively useful form of a de Finetti theorem is arguably the \emph{post-selection de Finetti theorem} of Christandl et al.~\cite{christandl2009postselection}, which uses Schur's lemma to express the projector on the symmetric subspace as a convex combination of tensor products. At the opposite end of the spectrum lies the \emph{exponential de Finetti theorem} of Renner~\cite{renner2008security}, which provides strong bounds but whose proof appears to be much more intricate. 

The purpose of this note is to give a simple, self-contained proof of the exponential de Finetti theorem. Our proof can be seen as a reduction, based on simple observations and a standard concentration argument (the Chernoff bound), from the exponential to the post-selection de Finetti theorem. The outline of the argument follows closely that of~\cite{renner2008security}, and we do not claim any originality; nevertheless we hope that our presentation may make the exponential de Finetti more accessible to some readers. 

We find it convenient to state the exponential de Finetti theorem in terms of a notion of \emph{quantum Hamming distance}. Let $\Hilb$ denote a Hilbert space, $\State(\Hilb)$  the set of pure states in $\Hilb$ and $\Density(\Hilb)$ the set of density matrices on $\Hilb$. Given an integer $n\geq 1$, $r\in\{0,\ldots,n\}$ and $\ket{\psi}\in\State(\Hilb^{\otimes n})$, the quantum Hamming ball of radius $r$ around $\ket{\psi}$ is defined as 
\begin{equation*}
	\Delta_r(\ket{\psi}) = \mathrm{span}\, \left \{ P \ket{\psi} : \text{ $P$ is a $r$-local unitary operator} \right \} \subseteq \State(\Hilb^{\otimes n}),
\end{equation*}
where by $r$-local we mean that $P$ is a unitary that acts as the identity on at least  $n-r$ subsystems of $\Hilb^{\otimes n}$. Given $\tau \in \Density(\Hilb^{\otimes n})$, its quantum Hamming distance to $\ket{\psi}$ is then defined as
\begin{equation*}
	\Delta(\tau,\ket{\psi}) = \text{minimum $r$ such that $\tau$ is supported on $\Delta_r(\ket{\psi})$.}
\end{equation*}

We are now ready to state (our variant of) the exponential de Finetti theorem:

\begin{theorem}[Exponential de Finetti theorem]\label{thm:df}
	Let $\Hilb = \C^d$ and $\rho \in \Density(\Hilb^{\otimes n+k})$ be a pure density matrix (i.e., a rank-$1$ state) invariant under permutation of the $n+k$ subsystems.  
Then for all $r\in\{0,\ldots,n\}$ and $\ket{\psi}\in\State(\Hilb)$ there is a  $\tau_\psi \in \Density(\Hilb^{\otimes n})$ such that $\Delta(\tau_\psi,\ket{\psi}^{\otimes n}) \leq r$ and
	\begin{align}
		\Big\| \Tr_k (\rho) - \int_{\ket{\psi} \in \State(\Hilb)} \tau_\psi \,\, \dif \nu(\psi) \Big\|_1 \leq (n+k)^{O(d)} \cdot \exp \Big ( - \frac{r}{6} \cdot \min \left \{ \frac{k}{n} , 1 \right \} \Big),
		\label{eq:df}
	\end{align}
	where $\Tr_k (\rho)$ denotes the partial trace with respect to the last $k$ subsystems and $\nu(\psi)$ is a measure on $\State(\Hilb)$ proportional to the Haar measure $\dif\psi$ weighted by $\Tr((\Id\otimes \ketbra{\psi}{\psi}^{\otimes k} )\rho)$. 
\end{theorem}

In~\cite{renner2008security} the bound obtained in the right-hand side of~\eqref{eq:df} is $2k^{O(d)} \exp \left( -\frac{(r+1)k}{2(n+k)} \right)$. For fixed $d$, ignoring the polynomial prefactors in front of the exponential and up to the exact constant in the exponent, the two bounds are equivalent since  $\min \{ k/n, 1 \} \leq 2k/(n + k) \leq 2\min \{ k/n, 1\}$.

As in~\cite{renner2008security}, Theorem~\ref{thm:df} can be extended to non-pure states by using the fact that any permutation-invariant density matrix, which may not be supported on the symmetric subspace,\footnote{We thank Renato Renner for pointing out this subtle issue, which the previous version of this note neglected.} can be purified to a symmetric rank-$1$ state (c.f. Lemma 4.2.2 in~\cite{renner2008security}). 

\section{Tools}

Our proof is based on three commonly used ingredients from probability theory and quantum information theory: the Chernoff bound, the Gentle Measurement lemma, and an integral representation of the projector on the symmetric subspace of $(\C^d)^{\otimes n}$. 

\begin{proposition}[Chernoff bound~\cite{chernoff1981note}]\label{prop:chernoff}
Let $X_1,\ldots,X_n$ be i.i.d. random variables taking values in $\{0,1\}$, and $\mu = \Ex[X_i]$. Then for all $0 < \alpha < 1$ such that $(1 + \alpha) \mu \leq 1$, we have
$$
	\Pr \Big( \frac{1}{n} \sum_{i=1}^n X_i > (1 + \alpha) \mu  \Big) \leq e^{- D((1+\alpha)\mu\|\mu)n},
$$
where $D(x\|y) = x\ln \frac{x}{y} + (1-x)\ln \frac{1-x}{1-y}$ is the relative entropy function. 
\end{proposition}

\begin{proposition}[Gentle Measurement lemma~\cite{ogawa2002new}]\label{prop:gentle}
For all positive semidefinite operators $\rho$, $X$ such that $0 \leq X \leq \Id$, 
$$
	\left \| \rho - \sqrt{X} \rho \sqrt{X} \right \|_1 \leq 2 \sqrt{\Tr( \rho)} \sqrt{\Tr \left ( \rho \left ( \Id - X \right) \right ) }.
$$
\end{proposition}
 
\begin{proposition}
\label{prop:postselection_df}
The projector $\Pi^{sym}_n$ on the symmetric subspace of $(\C^d)^{\otimes n}$ can be expressed as
$$
\Pi^{sym}_n= c_{n,d} \int_{\ket{\theta} \in \State(\Hilb)} \ketbra{\theta}{\theta}^{\otimes n} \, \dif \theta,
$$
where $c_{n,d}=\binom{n + d -1}{n}$ and $\dif \theta$ stands for the Haar measure on $\State(\Hilb)$. 
\end{proposition}

This last proposition can be proven using Schur's lemma; see e.g.~\cite{harrow2013church}.

\section{The proof}

 Let $\Pi^{sym}_{k}$ be the orthogonal projector on the symmetric subspace of $\Hilb^{\otimes k}$. Using the assumption that $\rho \in \Density(\Hilb^{\otimes n+k})$ is a pure symmetric state, tracing out the last $k$ registers yields
	\begin{align}
	\Tr_k(\rho) &=\Tr_k\big(\rho\, (\Id\otimes \Pi^{sym}_k )\big)\notag\\ 
&= 	c_{k,d} \int_{\ket{\psi} \in \State(\Hilb)} (\Id\otimes \bra{\psi}^{\otimes k}) \, \rho \, (\Id\otimes \ket{\psi}^{\otimes k}) \, \dif \psi,\label{eq:df-2}
\end{align}
	where the second equality is by Proposition~\ref{prop:postselection_df}. For any $\ket{\psi} \in \State(\Hilb)$ and $S \subseteq \{1,\ldots,n\}$ define
	$$P_{\psi,S} := \bigotimes_{i\notin S} \ket{\psi}\bra{\psi}_i \bigotimes_{i\in S} (\Id-\ket{\psi}\bra{\psi})_i,$$
	where a subscript $i$ indicates on which copy of $\Hilb$ each operator acts. For $S \neq S'$, $P_{\psi,S}P_{\psi,S'}=0$, so that for any $r\in\{0,\ldots,n\}$ we may define a pair of orthogonal projectors 
	$$P_{\psi}^{\geq r} = \sum_{S\subseteq \{1,\ldots,n\} :\, |S|\geq r} P_{\psi,S}\qquad\text{and}\qquad P_\psi^{< r} = \Id - P_\psi^{\geq r}.$$
		Let 
		$$\rho_\psi = (\Id\otimes \bra{\psi}^{\otimes k} )\, \rho \, (\Id\otimes \ket{\psi}^{\otimes k} ), \qquad \sigma_\psi = P_\psi^{< r} \, \rho_\psi \, P_\psi^{< r},\qquad \tau_\psi = \frac{1}{\Tr(\sigma_\psi)}\sigma_\psi,$$ 
		and note that $\Delta(\tau_\psi,\ket{\psi}^{\otimes n}) < r$. Applying the triangle inequality, Proposition~\ref{prop:gentle}, and Jensen's inequality, from~\eqref{eq:df-2} we obtain
\begin{align}
 \Big\| \Tr_k(\rho) -  c_{k,d} \int_{\ket{\psi}} \tau_\psi \,\, \Tr(\rho_\psi) \dif \psi \Big\|_1 &\leq \Big\| \Tr_k(\rho) -  c_{k,d} \int_{\ket{\psi}} \tau_\psi \,\, \Tr(\sigma_\psi) \dif \psi \Big\|_1 + c_{k,d} \int_{\ket{\psi}} \big|\Tr(\rho_\psi) - \Tr(\sigma_\psi)\big| \dif \psi \notag \\
 &\leq 3\, c_{k,d}\Big( \int_{\ket{\psi}} \Tr \left( P_{\psi}^{\geq r} \rho_\psi \right) \dif \psi \Big)^{1/2}.
 \label{eq:df-1}
\end{align}
Setting the measure $\dif\nu$ to be proportional to $\Tr(\rho_\psi)\dif \psi$ the theorem will be proved once we obtain an appropriate upper bound on the right-hand side of~\eqref{eq:df-1}. Using again the fact that $\rho$ is a pure symmetric state and Proposition~\ref{prop:postselection_df}, for any $\ket{\psi}\in\State(\Hilb)$ it holds that 
	\begin{equation}\label{eq:de_finetti}
	(\Id\otimes	\bra{\psi}^{\otimes k}) \, \rho \, (\Id\otimes \ket{\psi}^{\otimes k}) \leq c_{n+k,d} \int_{\ket{\theta} \in \State(\Hilb)} \ketbra{\theta}{\theta}^{\otimes n} \, | \braket{\theta}{\psi} |^{2k} \, \dif \theta.
	\end{equation}
	For $\ket{\psi},\ket{\theta}\in\State(\Hilb)$ let $x_{\theta\psi} := | \braket{\theta}{\psi} |^2$. From~\eqref{eq:de_finetti},	
	\begin{align}
			\Tr \left( P_{\psi}^{\geq r} (\Id\otimes \bra{\psi}^{\otimes k} )\,\rho\,(\Id\otimes \ket{\psi}^{\otimes k}) \right) &\leq c_{n+k,d}   \int_{\ket{\theta}} x_{\theta\psi}^k \, \sum_{|S| \geq r} x_{\theta\psi}^{n - |S|} (1 - x_{\theta\psi})^{|S|} \, \dif \theta \notag \\
		&= c_{n+k,d} \int_{\ket{\theta}} x_{\theta\psi}^k\, \sum_{i \geq r} \binom{n}{i} x_{\theta\psi}^{n - i} (1 - x_{\theta\psi})^{i} \, \dif \theta.
		\label{eq:definetti}
	\end{align}
	To bound the expression under the integral sign we distinguish two cases. If $x_{\theta\psi} \geq 1 - \frac{r}{3n}$, Proposition~\ref{prop:chernoff} applied to i.i.d. Bernoulli random variables $X_1,\ldots,X_n$ such that $\Pr(X_i=0) = x_{\theta\psi}$ shows that the binomial sum is at most  $\exp (-r/3)$. If $x_{\theta\psi} < 1 - \frac{r}{3n}$, then $x_{\theta\psi}^k \leq  \exp(- r k/3n)$. Taking the maximum gives an upper bound on the right-hand side of~\eqref{eq:df-1} and proves the theorem.

\bibliographystyle{plain}
\bibliography{df}

\end{document}